\documentclass{jpsj-suppl}
\usepackage{txfonts} 

\title{Equation of state for neutron star matter with NJL model and Dirac-Brueckner-Hartree-Fock approximation }

\author{Takahide Kambe, Tetsuya Katayama and Koichi Saito}

\inst{Department of Physics, Faculty of Science and Technology, Tokyo University of Science, Noda 278-8510, Japan}

\email{6215605@ed.tus.ac.jp}

\recdate{August 20, 2016}

\abst{In the present study, using the flavor-SU(3) NJL model with the vector coupling interaction, we have calculated the equation of state (EoS) for the quark phase at high density. Furthermore, for the hadron phase at low density, we have used two kinds of the equations of state; one is a rather soft one by the QHD model, and the other is a stiff one calculated with relativistic Brueckner-Hartree-Fock approximation [1]. Using those equations of state for the two phases, we have investigated the influence of various choices of parameters concerning the crossover region on the mass and radius of a neutron star.}

\kword{neutron star, hadron-quark crossover, Dirac-Brueckner-Hartree-Fock approximation, Nambu-Jona-Lasinio model.}

\begin{document}
\maketitle

\section{Introduction}

As the interior density of a neutron star becomes very high, it has been expected and discussed that quark matter may be created inside it. To describe the transition from hadron to quark phases (and vice versa), two mechanisms are mainly considered; one is the first-order phase transition [2], and the other is the crossover phenomenon [3, 4]. In the latter case, it is possible to explain the massive neutron stars using a quark-core. However, there is a considerable uncertainty in parameters and crossover mechanism. It is thus very vital to investigate the properties of crossover method carefully. 

\section{Crossover method}

To describe the crossover region between hadron and quark phases phenomenologically, we may need two parameters [3, 4]; the central density of the crossover region, $\bar{\rho}$,  and its width, $\Gamma$.  We here consider the following three cases: 
\begin{enumerate}
\item If we want to express the (total) energy density, $\epsilon$, using the energy densities of hadron and quark phases ($\epsilon_H$ and $\epsilon_Q$, respectively), as a function of baryon density, $\rho$, it may be given by an interpolation 
\begin{eqnarray}
\epsilon(\rho) = \epsilon_{H}(\rho)f_{-}(\rho) + \epsilon_{Q}(\rho)f_{+}(\rho)~,
\end{eqnarray}
with the following function 
\begin{eqnarray}
f_{\pm}=\frac{1}{2}\bigg[1\pm \tanh\bigg( \frac{\rho -\bar{\rho}}{\Gamma}\bigg) \bigg]~.
\end{eqnarray}
This choice is called the energy-density (E-D) interpolation [3].  
Using the thermodynamical consistency, pressure in this case is then calculated by 
\begin{eqnarray}
P(\rho) &=& P_{H}(\rho)f_{-}(\rho) + P_{Q}(\rho)f_{+}(\rho) + \Delta{P}(\rho)~,\\[5pt]
\Delta{P}(\rho)&=& \frac{2\rho}{\Gamma}\Big(\epsilon_{Q}(\rho) - \epsilon_{H}(\rho)\Big) \left( e^{(\rho-\bar{\rho})/\Gamma} + e^{-(\rho-\bar{\rho})/\Gamma} \right)^{-2}~,
\end{eqnarray}
where $P_{H}$ and $P_{Q}$ are respectively pressures in hadron and quark phases. 
\item  If we want to interpolate pressure as a function of the baryon density, it is calculated as 
\begin{eqnarray}
P(\rho) = P_{H}(\rho)f_{-}(\rho) + P_{Q}(\rho)f_{+}(\rho)~. 
\end{eqnarray}
In this case, the energy density is given by using the thermodynamical consistency 
\begin{eqnarray}
\epsilon(\rho)&=& \epsilon_{H}(\rho)f_{-}(\rho) + \epsilon_{Q}(\rho)f_{+}(\rho) + \Delta{\epsilon}(\rho)~,\\[5pt]
\Delta{\epsilon}(\rho) &=& \frac{2\rho}{\Gamma} \int_{\bar{\rho}}^{\rho} \frac{d{\rho^{'}}}{\rho^{'}}\Big(\epsilon_{H}(\rho^{'}) - \epsilon_{Q}(\rho^{'})\Big) \left(e^{(\rho^{'}-\bar{\rho})/\Gamma} + e^{-({\rho}^{'} - \bar{\rho})/\Gamma}\right)^{-2}~.
\end{eqnarray}
This is called the pressure-density (P-D) interpolation [3].
\item
We lastly consider the pressure-energy (P-E) interpolation, in which pressure is expressed as a function of the energy density as [4]  
\begin{eqnarray}
P(\epsilon) &=& P_{H}(\epsilon)f_{-}(\epsilon) + P_{Q}(\epsilon)f_{+}(\epsilon)~,\\[5pt] 
\rho &=& \rho_{0}\int_{\epsilon_{0}}^{\epsilon} \frac{d\epsilon^\prime}{P(\epsilon^\prime) + \epsilon^\prime}~.
\end{eqnarray}
\end{enumerate}

\section{Models of neutron star matter}

We have used two different models for hadron phase to study the properties of crossover region. The first is the result with DBHF (Dirac-Brueckner-Hartree-Fock approximation) [1] that includes the effects of Pauli exclusion, negative-energy states of baryons and short-range correlations between two baryons. This model can provide a rather hard EoS for neutron-star matter, and it can thus explain the observed neutron-star mass with more than 2 times the solar mass without any quark core. The other is the result with QHD (Quantum hadrodynamics) model including non-linear $\sigma$ terms in the Hartree level. Compared with the DBHF result, this model  provides a soft EoS, and the maximum mass of a neutron star becomes about 1.6 times the solar mass. Therefore, in this case, the effect of quark core inside a neutron star is necessary to reproduce the recently observed, maximum mass. 

In addition, we have calculated the EoS of quark phase by the flavor-SU(3) NJL (Nambu-Jona-Lasinio) model with the vector coupling interaction. We have adopted the Hatsuda-Kunihiro parameter set [5].

\section{Results}

\begin{figure}[t]
 \centering
    \includegraphics[width=15.0cm]{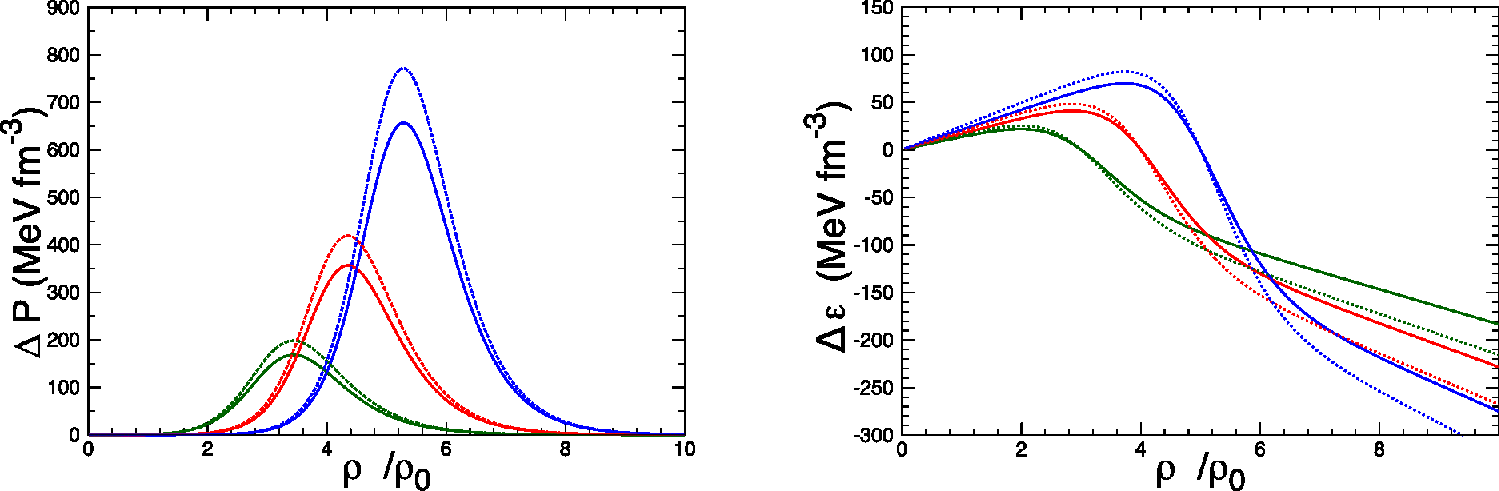}
\caption{$\Delta{P}$ (left) and $\Delta{\epsilon}$ (right) as a function of baryon number density in units of normal nuclear matter density, $\rho_{0}$= 0.17$\mathrm{fm}^{-3}$:  solid lines; DBHF model, dotted ones; QHD model. Each color designates:  green; $\bar{\rho}=3\rho_{0}$, red; $\bar{\rho}=4\rho_{0}$, blue; $\bar{\rho}=5\rho_{0}$. $\Gamma$ is fixed to be $\rho_{0}$, because the change of $\Gamma$ dose not give a large difference in the result.}
\label{fig:one}
\end{figure}

We show our results including the crossover effect between hadron and quark phases. In Fig. \ref{fig:one}, we present the corrections of pressure, $\Delta{P}$, and energy density, $\Delta\epsilon$, in the E-D and P-D interpolations, respectively.  

\begin{figure}[t]
  \centering
  \includegraphics[width=15cm]{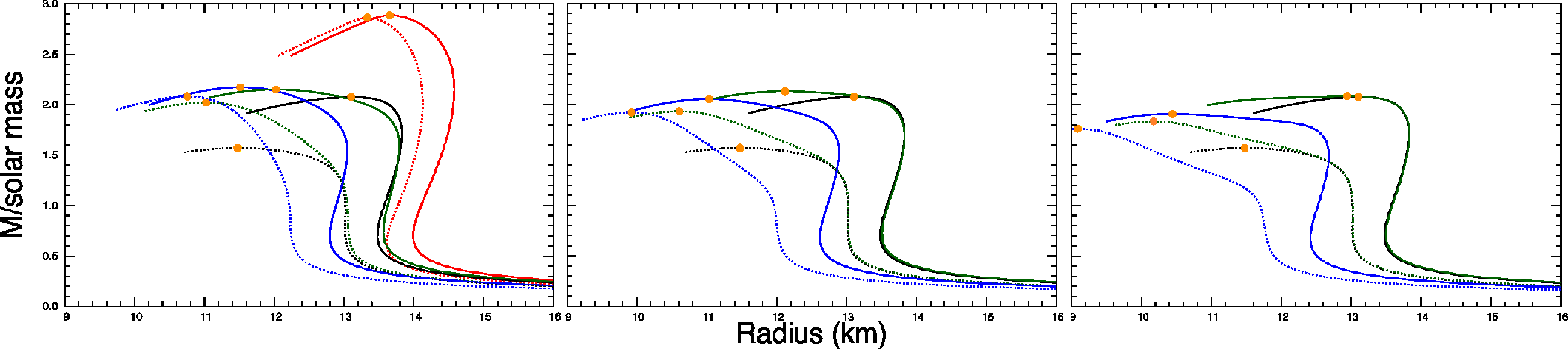}
  \caption{Mass-radius relation.  We choose the following parameters:  $g_{v}= G_{s}$ and $\Gamma = \rho_{0}$, and the left (middle) [right] panel is for $\bar{\rho} = 3\rho_{0} \, (4\rho_{0}) \, [5\rho_{0}]$. Each color shows: black; only hadron phase, red; E-D interpolation, green; P-D interpolation, blue; P-E interpolation. The dot on the line represents the position of the maximum mass.}
\label{stellation}
\end{figure}

In the E-D interpolation, $\Delta{P}$ stiffens the EoS in the entire density region. On the other hand, in the P-D interpolation, $\Delta\epsilon$  stiffens the EoS at only high-density. 
This correction leads a large impact on the radius and the maximum mass of a neutron star, which is shown in the mass-radius relation of neutron star (see Fig. \ref{stellation}).  In case of $\bar{\rho} \geq 4\rho_{0}$, because the causality is violated in the E-D interpolation scheme, we do not show the calculated results (see the middle and right panels). We have found that the E-D interpolation (red lines in Fig. \ref{stellation}) gives a large mass and radius because of $\Delta{P}$. In contrast, the P-D interpolation provides a small radius due to the $\Delta\epsilon$ correction (see blue lines).   

On the other hand, in the P-E interpolation scheme, the results (green lines) are very close to those of hadron phase only (black lines). Furthermore, we can see the dependence of the crossover central density $\bar{\rho}$ on the EoS: as $\bar{\rho}$ becomes high, the P-E results approach the EoS of hadron phase only. However, the P-D results become away from the results of hadron phase only. 

\newpage
\section{Conclusion}

In this study, we have calculated the properties of neutron stars by choosing various values of $\bar{\rho}$ and $\Gamma$ and interpolation mechanism.  We then have found the strong dependence of the parameters and interpolation mechanism on the radius and the maximum mass of a neutron star.  In the future, if the radius of a massive neutron star could be observed with high accuracy, it would be able to determine which mechanism and parameters are more favorable in the crossover method.


\end{document}